# A high-speed, high-resolution Transition Edge Sensor spectrometer for soft X-rays at the Advanced Photon Source

Orlando Quaranta, Don Jensen, Kelsey Morgan, Joel C. Weber, Jessica L. McChesney, Hao Zheng, Tejas Guruswamy, Jonathan Baldwin, Ben Mates, Nathan Ortiz, Johnathon Gard, Doug Bennet, Dan Schmidt, Lisa Gades and Antonino Miceli

*Abstract*— This project explores the design and development of a transition edge sensor (TES) spectrometer for resonant soft X-ray scattering (RSXS) measurements developed in collaboration between Argonne National Laboratory (ANL) and the National Institute of Standards and Technology (NIST). Soft X-ray scattering is a powerful technique for studying the electronic and magnetic properties of materials on a microscopic level. However, the lack of high-performance soft X-ray spectrometers has limited the potential of this technique. TES spectrometers have the potential to overcome these limitations due to their high energy resolution, high efficiency, and broad energy range. This project aims to optimize the design of a TES spectrometer for RSXS measurements and more generally soft X-ray spectroscopy at the Advanced Photon Source (APS) 29-ID, leading to improved understanding of advanced materials.

We will present a detailed description of the instrument design and implementation. The spectrometer consists of a large array of approximately 250 high-speed and high-resolution pixels. The pixels have saturation energies of approximately 1 keV, sub-ms pulse duration and energy resolution of approximately 1 eV. The array is read out using microwave multiplexing chips with MHz bandwidth per channel, enabling efficient data throughput. To facilitate measurement of samples in situ under ultra-high vacuum conditions at the beamline, the spectrometer is integrated with an approximately 1 m long snout.

*Index Terms*—Beamline, Cryogenic, Detector, Resonant Soft X-ray Scattering, Spectroscopy, Superconductivity, Transition Edge Sensor.

## I. Introduction

TRANSITION EDGE SENSORS (TESs) are emerging as revolutionary tools in the field of radiation detection, particularly in the application of X-ray spectroscopy at synchrotron facilities [1], [2], [3], [4], [5].

These superconducting devices offer unprecedented sensitivity and resolution, characteristics that are critical for the cutting-edge experiments conducted at synchrotron light sources for probing subtle changes in electronic and magnetic structures. This article will describe a new soft X-ray instrument in development at the Advanced Photon Source (APS) of Argonne National Laboratory for the beamline 29-ID. Specifically, this work will provide detail on the sensor design, readout electronics and cryogenic with necessary integration with the existing beamline instrumentation. We will also examine how such an instrument will enhance the capabilities of Resonant Soft X-ray Scattering (RSXS) studies, enabling researchers to gain deeper insights into complex material behaviors and interactions at the atomic level. By pushing the boundaries of what is possible in soft X-ray spectroscopy, TESs are instrumental in driving new discoveries and innovations in fields ranging from condensed matter physics to chemical dynamics.

## II. TES Application to Resonant Soft X-ray Scattering

Resonant soft x-ray scattering (RSXS) has become a paramount technique for exploring valence-band order in quantum materials, notably identifying charge-density-wave (CDW) order across various copper-oxide superconductors [6]. RSXS utilizes a quasi-elastic scattering method that leverages resonance between core and valence levels, enhancing valence-band sensitivity. This technique captures an electron spectral function, akin to what is observed in STM experiments, by involving intermediate states consisting of a core hole coupled with an extra valence electron. Despite its advantages, the effectiveness of RSXS is often thwarted by overwhelming photoabsorption, which increases the incoherent radiative background and is particularly problematic when dealing with diffuse scattering structures (typically within a few eV of the coherently scattered light).

To mitigate this issue, using energy-resolving detectors like Transition Edge Sensors offers a promising solution. Previously successful in various spectroscopic applications [1], [2], [3], [4], TES detectors provide exceptional energy resolution and quantum efficiency which outperform traditional solid-state detectors in resolving energies. Their enhanced energy resolution, together with their very low background, allows for

"This research was funded by Argonne National Laboratory LDRD proposals 2018-002 and 2021-0059; is supported by the Accelerator and Detector R&D program in Basic Energy Sciences' Scientific User Facilities (SUF) Division at the Department of Energy; used resources of the Advanced Photon Source and Center for Nanoscale Materials, U.S. Department of Energy (DOE) Office of Science User Facilities operated for the U.S. DOE, Office of Basic Energy Sciences under Contract No. DE-AC02-06CH11357". (Corresponding author: Orlando Quaranta).

O. Quaranta (e-mail: oquaranta@anl.gov), D. Jensen, Jessica L. McChesney, H. Zheng, T. Guruswamy, J. Baldwin, L. Gades and A. Miceli are with Argonne National Laboratory, Lemont, IL 60439 USA.

K. Morgan, J. C. Weber, B. Mates, N. Ortiz, J. Gard, D. Bennet, and D. Schmidt are with the National Institute of Standards and Technology, Boulder, CO 80305 USA.

ASC2024-1EOr2D-042

the discrimination necessary in RSXS, potentially enhancing the detection of subtle signals [6]. Recent advancements have also facilitated their integration into large-scale x-ray experiments [7], significantly increasing data collection efficiency through simultaneous analysis at multiple reciprocal space points. This development marks a substantial improvement over traditional methods, streamlining RSXS applications in studying complex materials.

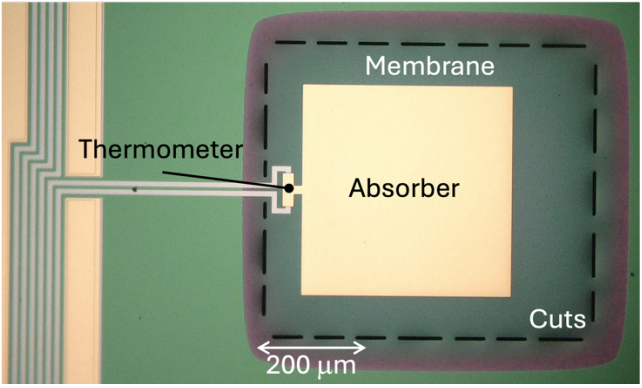

**Fig. 1.** Optical microscope photo of a representative TES.

### III. PIXEL DESIGN

A series of test devices have been designed, fabricated and characterized to identify the one best suited to the specifications needed for use at APS-29-ID, in the form of 24-pixel dies. The specification can be summarized as following: Saturation Energy ($E_{SAT}$) ~ 1 keV, Energy Resolution ($\Delta E$) ≤ 1, Pulse Duration ≤ 1 ms and Critical Temperature ($T_C$) ≥ 90 mK.

These needs have been defined as a combination of various competing factors. In RSXS experiments the resonant signal can be overwhelmed by the emission of core level fluorescence photons, which can be within few eV of the coherently scatter light, this set a minimum boundary to the energy resolution [7]. Although better energy resolutions could be archived by operating a very low temperatures (50 mK) [8], this would require the use of a Dilution Fridge, which is incompatible with the environment at APS-29-ID. Therefore, an operational temperature of approximately 100 mK was chosen as a compromise between archivable energy resolution and cryostat specifications. Finally, the pulse duration is set to be as fast as possible, while still be compatible with the multiplexing readout scheme described below. Fast pulses are needed to make the overall acquisition time (over all pixels in the array) in tune with the typical experiment duration at a beamline.

The devices are all based on a "sidecar" design, in which the superconducting thermometer is placed alongside a photon absorber, and thermally connected to it via a metallic patch. Both the thermometer and the absorber sit on a suspended membrane, which provides the thermal connection to the cryogenic environment. The thermometer is composed of an Au/Mo bilayer, with the Mo layer (also defining the wiring) deposited via sputtering and subsequent etch, and the Au layer deposited via e-beam evaporation and subsequent liftoff. The absorber is composed of Au over a thin adhesion layer of Ti (2 nm), deposited via e-beam evaporation and subsequent lift-off. Finally, the membrane is fabricated from SiN, released by a backside DRIE etch to a SiN/Si wafer, together with eventual cuts to more accurately define the device thermal conductance ($G$). The overall dimensions of the device parts are reported in Table I, while a photo of a representative design is shown in Fig. 1. The various designs were present in each die in multiple copies. To bias the devices a shunt resistance of $R_{SH}$ = 250 µΩ and an inductance of $L$ = 140 nH were used. The TESs were readout via the use of a microwave multiplexing chips (µmux) [9]. The µmux chips are composed of 63 resonators, each with a bandwidth of 1 MHz and spaced by 7 MHz (for a total bandwidth of ~ 2 GHz) and typical noise levels of 2-3 µΦ$_0$/√Hz.

TABLE I
TES DIMENSIONS

|  | Thermometer (µm) | Absorber (µm) | Membrane (µm) |
|---|---|---|---|
| Length ($l$) | 90 | 250 – 400 | 750 |
| Width ($w$) | 30 | 250 – 400 | 750 |
| Thickness ($t$) | Au/Mo 0.500/0.045 | 0.100 – 0.250 | 0.500 |

The various device designs varied only the dimensions of the X-ray absorber, to explore the interplay between heat capacity ($C$), saturation energy, energy resolution and position dependent photon response.

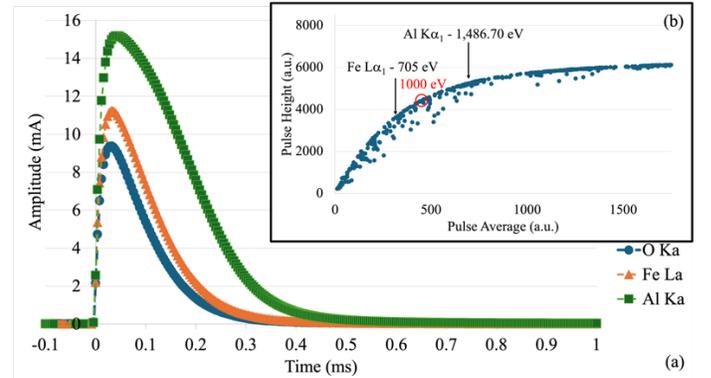

**Fig. 2.** (a) Candidate TES ($w$ = 250 µm, $l$ = 250 µm and $t$ = 0.150 µm) pulses for O Kα$_1$ (524.9 eV), Fe Lα$_1$ (705 eV) and Al Kα$_1$ (1,486.70 eV) photons and (b) linearity curve.

Of all the various combinations of the absorber lateral dimensions and/or thicknesses, after averaging multiple device measurements, the one that proved to perform best was the following: $w$ = 250 µm, $l$ = 250 µm and $t$ = 0.150 µm. Devices characterized by larger absorbers (either in lateral dimensions or thickness), although more efficient in detecting photons, were characterized by worse energy resolutions. For devices with thinner absorbers, the resolution didn't improve as much as expected (a possible indication of position dependence effects). The improvement in resolution did not justify the loss in efficiency.

In Fig. 2. (a) are shown examples of pulses collected by these detectors for photons of increasing energy (O K$\alpha_1$ – 524.9 eV, Fe L$\alpha_1$ – 705 eV and Al K$\alpha_1$ – 1,486.70 eV), while Fig. 2. (b) is showing the linearity of the device response (by comparing the pulse height to the pulse average). The performances of these kind of devices, as averaged over multiple identical devices present on the same chip, are the following: $E_{SAT}$ > 1 keV, $\Delta E$ = 2.09 ± 0.11 eV (statistical uncertainty), Fall Time ($\tau_{Fall}$) ~ 0.0804 ms, Thermal Conductance ($G$) ~ 120 pW/K and $T_C$ ~ 105 mK. The energy resolution was measured at the Al K$\alpha_1$ line and was measured by deconvolving the device spectral response with the intrinsic line shape of Al K$\alpha_1$ from [10]. The expected resolution from the device noise (integrated noise equivalent power) at this energy was 1.48 eV, while at Fe L$\alpha_1$ was 1.25 eV. The lack of detailed information on the intrinsic line width of Fe L$\alpha_1$ prevents us from estimating the device resolution for this line.

The performance matches the requirements from the pulse duration, saturation energy and operational temperature point of view, but fails to reach the wanted energy resolution. Based on these results, it has been decided that the final array for the instrument will be based on the design reported above but aiming at a lower critical temperature ($T_C$) of approximately 90 mK, still compatible with ADR operation. This should bring the expected energy resolution to ~1.07 eV at Fe L$\alpha_1$ and 1.29 eV at Al K$\alpha_1$, closer to the targeted 1 eV, while maintaining linearity approximately up to 1 keV. The final design will also include an extra layer of Bi of 2 to 3 μm (there is some variation across the wafer) in the absorber to further increase the detection efficiency [11], [12], [13]. This will add a negligible amount to the total heat capacity, which will not affect the energy resolution. The final array will be comprised of ~250 pixels arranged in a pattern as shown in Fig. 3. Here the devices are represented by the following color scheme: light blue is Mo, yellow is Au, black is Bi, light green is solid Si and in dark green suspended SiN membrane.

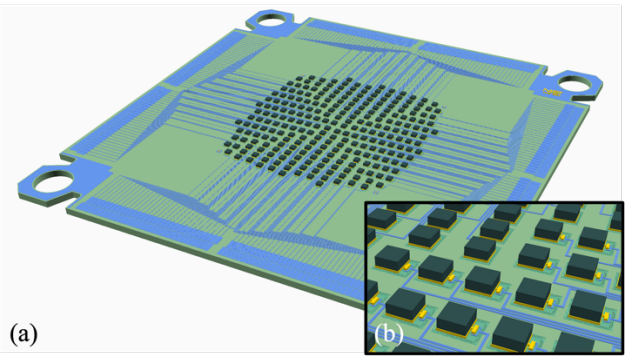

**Fig. 3.** (a) Rendering of the final ~250-pixel array. (b) close-up of the array where the single pixel design is visible. In light blue is Mo, in yellow Au and in black Bi. In light green is solid Si and in dark green suspended SiN membrane.

### IV. SENSORS READOUT

The final array will be integrated in a microsnout [14], together with necessary readout chips (four of each): shunt resistors ($R_{SH}$), Nyquist inductances ($L$) and microwave multiplexing chips (μmux) [15], as shown in Fig. 4. The μmux chips are of the same kind described in section III. These allow the correct sampling of pulses of the speed reported in the previous section for a total of 244 pixels (not all the resonators on a chip can be connected to a TES).

The μmux chips will be readout at room temperature via the use of Xilinx's Radio Frequency System on Chip (RFSoC) Field Programmable Gate Arrays (FPGAs) [16]. This FPGA family integrates:

- Digital to analog (DAC) and analog to digital (ADC) converters operating at radio frequency (RF) rates (up to 10 Gsample/s and 5 Gsample/s, respectively).
- Programmable logic FPGA, for high-speed Digital Signal Processing.
- Multicore ARM CPU subsystem with peripherals, capable of running an embedded Linux, control, system integration, and data transmission.

This level of integration greatly simplifies the overall system complexity, granting a sizable per-channel reduction in power consumption and cost with respect to ROACH2 systems that have been widely used in the past [17]. One Xilinx RFSoC system has a total readout bandwidth of 4 GHz, which allows for the readout of 2 complete microsnouts.

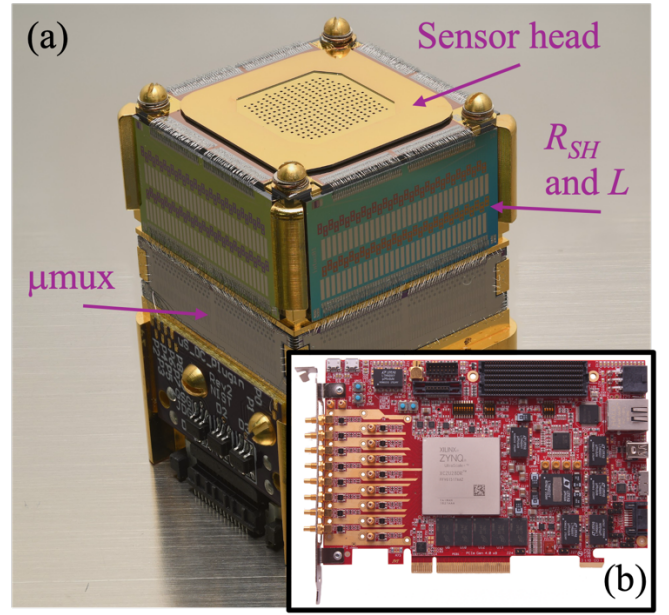

**Fig. 4.** (a) Photo of a representative microsnout. (b) High-Tech Global ZRF8 FPGA card.

### V. CRYOGENICS AND BEAMLINE INTEGRATION

This instrument will be installed into an existing beamline, APS 29-ID. This requires the ability to integrate it with existing systems, without interfering with the normal operation of the beamline. 29-ID is a soft X-ray beamline (photon energies up to 3 keV). This implies that all the experiments need to be performed inside a UHV environment to minimize the scattering from air. 29-ID has UHV chamber of ~ 1 m diameter with a goniometer inside and multiple access ports that allow the insertion of various instruments and a cryogenic sample stage. The sample stage is located at the center of the chamber. To effectively integrate the TES instrument with the chamber, a snout with the sensor head at the end of it was designed. To achieve the maximum level of compatibility, while still maintain high detection efficiency, a series of requirements need to be considered while designing the snout:

- The snout needs to extend the cryogenic environment present in the cryostat, while minimizing negative effect on the cryostat performance, including the presence of the necessary DC and RF wiring.
- High level of thermal, electrical and magnetic shielding of the sensor.
- Windows that can vacuum and thermally shield the environment in the snout/cryostat, while still allowing the penetration of soft X-rays.
- The chamber experiences vacuum levels much higher than that typical of cryostat. Moreover, the instrument needs to be completely independent from the beamline chamber, to allow for the decoupling of the two. Therefore, the two vacuum systems need to be completely decoupled.
- Ability to bring the sensor head as close as possible to the sample under study. This implies the ability to insert and retract the instrument as needed.

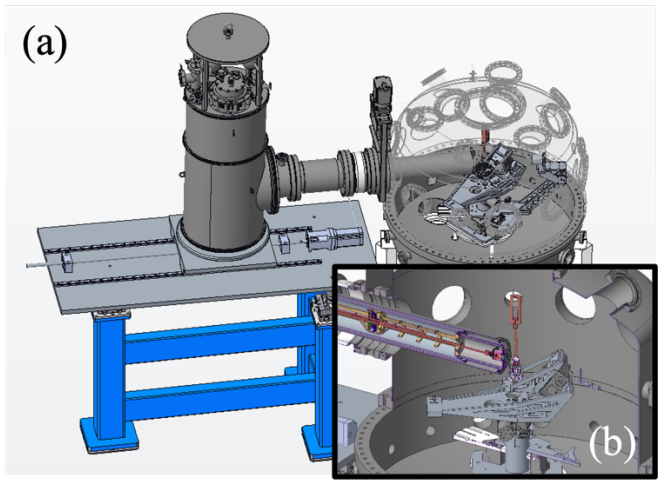

**Fig. 5.** (a) Rendering of the instrument (cryostat and snout) integrated with the beamline UHV chamber (transparent). (b) Cross-section of the beamline chamber and snout in proximity of the sample.

A system that meets all these requirements has been designed and is in construction. In Fig. 5 a rendering of the instrument integrated in the beamline is represented, with the chamber in transparency, where the various access ports, to which other instruments could be attached, and the goniometer are visible. This is the configuration with the instrument completely inserted in the chamber for maximum detection efficiency.

Details on the snout design are represented in Fig. 6. The main component of the snout (wand) is a frame that connects to the 3 K stage of the cryostat, that hosts two double-stage cryogenic pucks, which allows to extend the two ADR stages present in the cryostat (1 K and 50 mK) throughout the length of the snout. The microsnout sensor head is connected at the end of the 50 mK rod. The frame also reroutes the superconducting wiring (both DC and RF) from within the cryostat to the microsnout, while keeping it at 50 mK (the last anchor point is inside the main body of the cryostat). At the end of the wand, at 3 K, a magnetic shield in the form of a can of amumetal [18] around the microsnout is also present. The wand is completely self-supporting, allowing it to be completely disconnect from the main body of the cryostat without further support, which greatly facilitates maintenance far from the beamline. Around the wand a series of thermal shields are present at each cryostat's temperature stage (3 K, 60 K and 300 K), that connect to the respective shields on the main body of the cryostat. At the end of each thermal shield is also present a Luxel Al window [19] to allow X-rays in with minimum attenuation. The 300 K shield also acts as vacuum shield for the cryostat. Further, a vacuum bellows is also present around the 300 K shield, together with a vacuum valve and pumping port. This allows for the seamless insertion and extraction of the snout inside of the beamline vacuum chamber, while maintaining the vacuum levels of the chamber.

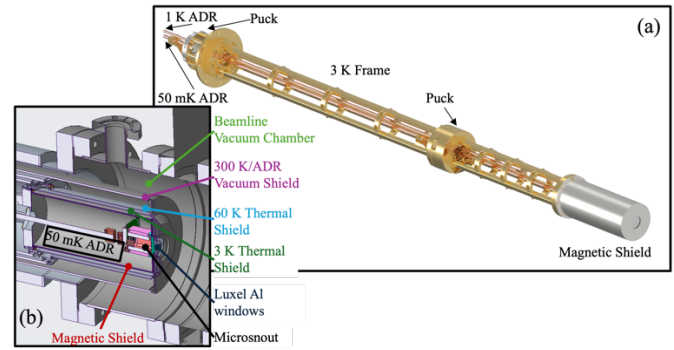

**Fig. 6.** (a) Rendering of the wand. (b) Cross-section of the sensor head with all the various thermal and vacuum shields visible.

## VI. CONCLUSIONS

In this work the recent developments towards the implementation of a high-speed, high-resolution Transition Edge Sensor spectrometer for soft X-ray for APS 29-ID have been presented. Details on the sensor design and performance, together with readout and integration solutions have been discussed. The instrument will be able to achieve sub-ms detection speed per pixel, with resolutions of ~1 eV, over an array of 244 pixels. These specifications will allow to run high-resolution, low-background RSXS experiments to study electron order parameters of complex materials, which will allow the study of complex phenomena like the interplay between superconductivity and other electronic order phenomena [6].